\documentclass[prl,twocolumn,showpacs]{revtex4}
\usepackage{epsfig}
\usepackage{graphicx}
\usepackage{amsmath}
\usepackage{mathrsfs}
\usepackage{dcolumn}
\newcommand{\beq}{\begin{eqnarray}}
\newcommand{\eeq}{\end{eqnarray}}
\newcommand{\be}{\begin{equation}}
\newcommand{\ee}{\end{equation}}
\newcommand{\bey}{\begin{eqnarray}}
\newcommand{\eey}{\end{eqnarray}}
\newcommand{\ba}{\begin{array}}
\newcommand{\ea}{\end{array}}
\newcommand{\bi}{\begin{itemize}}
\newcommand{\ei}{\end{itemize}}
\newcommand{\bem}{\begin{enumerate}}
\newcommand{\eem}{\end{enumerate}}
\newcommand{\bw}{\begin{widetext}}
\newcommand{\ew}{\end{widetext}}
\newcommand{\ra}{\rangle}
\newcommand{\la}{\langle}

\newcommand{\ov}{\overline}

\newcommand{\ww}{\widetilde}

\newcommand{\E}{{\cal E}}

\newcommand{\N}{{\cal{N} }}

\begin{document}

 \title{
 Preferred States of Decoherence under Intermediate System-Environment Coupling}
 \author{Wen-ge Wang$^{1,2}$ \footnote{ Email address: wgwang@ustc.edu.cn},
 Lewei He$^{1}$, and Jiangbin Gong$^{2,3}$ \footnote{ Email address: phygj@nus.edu.sg}}
 \affiliation{
 $^{1}$Department of Modern Physics, University of Science and Technology of China,
 Hefei, 230026, China
 \\ $^2$Department of Physics and Centre for Computational Science and Engineering,
 National University of Singapore, Singapore 117542 \\
$^{3}$ NUS Graduate School for Integrative Sciences
and Engineering, Singapore 117597
}

 \date{\today}

 \begin{abstract}
 The notion that decoherence rapidly reduces a superposition state to an incoherent mixture
 implicitly adopts
a special representation, namely, the representation of preferred (pointer) states (PS).
For  weak or  strong system-environment coupling, the behavior of PS is well known.
Via a simple dynamical model that simulates a two-level system interacting with few other
degrees of freedom as its environment,
it is shown that even for intermediate system-environment coupling, approximate PS may
still emerge from the coherent quantum dynamics of the whole system in the absence of any
thermal averaging. The found PS can also continuously deform to expected limits for weak
or strong system-environment coupling.
Computational results are also qualitatively explained.
The findings should be useful towards further understanding of decoherence and quantum
thermalization processes.
 \end{abstract}
\date{\today}
 \maketitle

 \emph{Introduction} --- As illustrated by the Schr\"{o}dinger cat paradox, there is a
 clash between the quantum
 superposition principle and the way we perceive the macroscopic reality.  So how does
 a classical world
 emerge from the quantum?  One promising solution to this profound question is
 {  decoherence}, i.e., the loss of quantum coherence due to the interaction of a system
 of interest with its environment~\cite{Zurek03,JZKGKS03,Schloss04}.  Decoherence may rapidly
 reduce a coherent superposition state of the system to an incoherent mixture. During this
 process the environment singles out special basis states, often called ``preferred (pointer)
 states" (PS), of which a classical probabilistic description becomes sufficient to describe
 the system
 and the bizarre superposition state of the PS is out of the picture.  That is, in the PS
 representation the reduced density matrix (RDM) of the system
 becomes diagonal as time evolves.

Such a decoherence perspective is not expected to resolve all conceptual issues
regarding quantum weirdness vs classical reality.
Nevertheless, it is highly useful as it implies the environmental-dependence of
the quantum-classical transition and the
representation-dependent nature of decoherence issues. For example, different environments
may select different PS, and a quantum system decohered in one PS representation may
still possess certain quantum coherence in other representations.
Going further, one may envision the possibilities of environment engineering to form
desired PS~\cite{viola}, such that system properties are robust to decoherence.

PS have been identified in several cases. If the system-environment
interaction is in the adiabatic limit or if it commutes with the
system's self-Hamiltonian, then populations on the energy
eigenstates of the system's self-Hamiltonian do not change but their
relative phases are destroyed by the environment. The energy
eigenstates then form the PS
\cite{Zurek81,BHS01,DK00,PZ99,GPSS04,ZHP93}. Analogous to this, if
the system-environment coupling is sufficiently weak, then the
energy relaxation time scale is typically much longer than the
pure-dephasing time scale. As a result, the energy eigenstates still form
the PS before relaxation sets in  \cite{WGCL08,Gogolin,Berman}.
Again, for weak system-environment interaction but for a longer time
scale, a model of quantum Brownian motion reveals that coherent
states localized in both position and momentum turn out to be the PS
\cite{Zurek03,JZKGKS03,Eisert04,ZHP93}. In the opposite situation,
the system-environment coupling is strong and the system's
self-Hamiltonian becomes negligible within a certain time period. In this
case, the eigenstates of the system-environment interaction
Hamiltonian form the approximate PS \cite{Zurek81,BHS01,PZ99}.

Little has been said about the possible existence of PS for a generic system-environment coupling
(i.e., not commutable with system's self-Hamiltonian) of
intermediate strength.  Under such a situation, the widely used quantum-master-equation approach
or other perturbative approaches may not be applicable in analyzing the existence of PS.
Also motivated by the ongoing investigations of quantum
thermalization processes~\cite{thermalNature,book},
we choose to work with a simple dynamical model to address the issue of PS. That is,  within a single
isolated quantum system composed
of interacting quantum subsystems, will the concept of PS still work well in describing
the decoherence of
one subsystem due to its interaction with other subsystems~\cite{gongpra03}?

We start from a computationally intuitive definition of PS.
We then show interesting evidence that PS may still exist for intermediate system-environment
coupling and further explain why this is possible.
The found PS, neither the system's energy eigenstates nor the eigenstates of the
system-environment interaction Hamiltonian,
undergo continuous deformation as the system-environment coupling strength varies.
These findings show that decoherence-induced superselection rule can be twofold:
superposition states of the PS are destroyed but PS themselves can be rich superposition states
for intermediate system-environment coupling.  Equally interesting, it can be concluded that
the concept of PS is still important in understanding quantum dynamical processes in the absence
of any thermal averaging.

\emph{Identifying PS from the time-evolving RDM} --- If PS exists, then the RDM
will gradually become diagonal in the PS representation.
On the other hand, the same RDM is always diagonal in its own eigenrepresentation. Therefore,
if we computationally track the eigenstates of the RDM, then we can see clearly whether or not
a well-defined set of PS can emerge
from a decoherence process. That is, if after a certain
period the eigenstates of the RDM
are found to evolve closely around a fixed basis set, then this fixed set of states can be
defined as the PS, at least approximately.
This computational definition of PS extends the PS criterion used by Di\'{o}si and Kiefer~\cite{DK00}.
Note also that such a definition
of PS gives up their precise analytical form.
Consistent with this picture, the off-diagonal elements of the RDM in the PS representation must
be also small when compared with
the difference of its diagonal elements. Indeed, were the RDM diagonal elements almost degenerate,
then a small fluctuation in the
off-diagonal elements can still cause a drastic rotation of the RDM eigenstates, a fact that would
contradict with
the existence of PS~\cite{epaps}.  We hence mainly work in
the parameter regimes where an appreciable difference between the diagonal elements of RDM can
emerge from the dynamics. These preliminaries
also make it clear that even a stationary RDM does not necessarily mean the existence of PS.

Consider now a two-level system $S$
 interacting with its environment $\E$, with a total Hamiltonian
 $ H = H_S + H_I + H_{\E}$,
 where $H_S$ and $H_{\E}$ are the Hamiltonians of $S$ and $\E$, and
 $ H_I$ is the system-environment interaction Hamiltonian, with $[H_{S}, H_{I}]\ne 0$.
 Eigenstates of $H_S$ are denoted by $|E_k\ra $, with $k=0,1$.
Throughout, we use $|\Psi\ra $ to denote a state vector for the whole system-environment
combination, denoted by $S+\E$ and
isolated from any thermal bath.
 The time-evolving RDM for $S$ is given by $\rho^s(t) \equiv
 {\rm Tr}_{\E} \left ( |\Psi(t)\ra \la \Psi(t)| \right )$, where $|\Psi(t)\ra
 = e^{-iHt/ \hbar } |\Psi(0)\ra $ coherently evolves according to Schr\"{o}dinger equation for  $S+\E$.

Eigenstates of the RDM $\rho^s(t)$ are represented by $|\rho_k(t) \ra$, with eigenvalues
$\rho_k(t)$, i.e., $ \rho^s(t) | \rho_k(t)\ra = \rho_k(t) | \rho_k(t) \ra,$ with $k=0,1$.
The two states $| \rho_0(t) \ra$ and $| \rho_1(t) \ra$ also form an orthonormal basis set for
the Hilbert subspace associated with $H_S$.
The distance between this basis set and another basis set  $|{\eta}_{k'}\rangle$ ($k'=0,1$)
for the same subspace
may be measured by $D(|\rho_k(t)\ra, |{\eta}_{k'}\ra ) =1-| \la\rho_k(t)|{\eta}_{k'}\ra |^2$,
with $k$ and $k'$ determined by the condition $|\la\rho_k|{\eta}_{k'}\rangle |^2 \ge 1/2$.
A time-averaged distance $d$ over a period $[t_a, t_b]$ can then be defined as
$d(|{\eta}_{k'}\ra )= [1/(t_b-t_a)] \int_{t_a}^{t_b} dt'\ D(|\rho_k(t')\ra, |{\eta}_{k'}\ra )$.
If, for a particular basis set $\{|\ww{\rho}_{k'}\ra \}$ defined below, $d(|\ww{\rho}_{k'}\ra)$ is small
for sufficiently large $t_a$, $t_b$, and $t_b-t_a$,
then $\rho^s(t)$ becomes almost diagonal in the $\{|\ww{\rho}_0\ra$, $|\ww{\rho}_1\ra \}$
representation, and hence $\{|\ww{\rho}_0\ra$, $|\ww{\rho}_1\ra\}$ can be
computationally identified as the PS.

To find $\{|\ww{\rho}_0\ra$, $|\ww{\rho}_1\ra \}$ that may reflect the
average behavior of $|\rho_k(t)\ra$ with acceptable fluctuations,  we calculate
the time-evolving RDM eigenstates $|\rho_k(t)\ra$, average the density matrix
$|\rho_{k}(t)\ra\langle\rho_k(t)|$ over time (value of $k=0$ or $k=1$ is chosen to maintain a
continuity),  and then obtain a time-averaged density matrix
${\overline{\rho}}$. Finally, the eigenstates of ${\overline{\rho}}$
are defined as the basis states $\{|\ww{\rho}_{0}\rangle , |\ww{\rho}_{1}\rangle\}$~\cite{notes}.
If, in the $\{|\ww{\rho}_{0}\rangle , |\ww{\rho}_{1}\rangle\}$ representation,
$|\rho^{s}_{01}(t)|\ll |\rho^{s}_{00}(t)-\rho^{s}_{11}(t)| $ for sufficiently large times $t$,
then $d(|\ww{\rho}_{k'}\ra)$
is small and PS can hence be identified~\cite{notes}.  If this is not the case, then PS fails to emerge
from the dynamics.

\emph{Model} --- We now turn to a concrete model.
 To reflect the fact that typically a small system $S$ is not directly coupled to the whole of its
  environment $\E$,
 we let $S$ be directly coupled to a small component $A$ of $\E$, and then
 let $A$ be further coupled to the rest part $B$ of $\E$, with $\E=A+B$.
 For convenience, $A$ is also assumed to be a two-level system. Such kind of coupling scheme
was also considered recently~\cite{italygroup} to model a nonlinear system-environment coupling.
It can be
also qualitatively argued that our coupling scheme can yield much less fluctuation
in the RDM than a full coupling between $S$ and $\E$ does.
The $B$ part of $\E$ is simulated by a quantum kicked rotor on a torus with only 1 degree of freedom,
whose classical limit is fully chaotic~\cite{RCB06,quantumchaosmodels}.
The irregular motion of $B$ due to quantum chaos, instead of
many noninteracting degrees of freedom of a thermal bath, is responsible for decoherence in $S$.
In terms of standard Pauli matrices and operators for a kicked rotor in dimensionless units,
the Hamiltonians for the system, the environment, and their coupling are
 \begin{eqnarray}
H_{S}&=&\omega_x \sigma_{x}^{S}+\omega_z\sigma_{z}^{S}, \  H_I =  \varepsilon\sigma^{S}_{z}
\otimes\sigma^{A}_{z},\nonumber \\
H_{\E}& =& H_{A}+H_{B}+H_{AB}.
\end{eqnarray}
 Here, $H_{A}=\omega_A\sigma_{x}^{A}$,
 $H_{B}=\frac{p^2}{2}+v\cos\gamma \sum_{j} \delta(t-jT)$,
and $H_{AB}=\lambda\sigma^{A}_{z}\cos\gamma\sum_{j}\delta(t-jT)$,
 where $p$ and $\gamma$ are momentum and coordinate operators of the kicked rotor.
 Since the system-environment coupling is already of the $\sigma_z^{S}$ type, for generality
 $H_S$ is made to contain
both $\sigma_{x}^{S}$ and $\sigma_{z}^{S}$ terms \cite{note}.  The unitary propagator associated with one
period $T$ is (with $\hbar =1$).
 \bey
 \hat{U}_T = e^{-iT(\omega_x{\sigma}^{S}_x+\omega_z{\sigma}^{S}_z
 +\omega_A{\sigma}^{A}_x+
 \varepsilon {\sigma}^{S}_{z}\otimes{\sigma}^{A}_{z})}\nonumber
 \\ \times e^{-iT\frac{{p}^2}{2}}e^{-iv\cos{\gamma}}
 e^{-i\lambda{\sigma}^{A}_{z}\cos{\gamma}}.
 \eey
The initial state is chosen as $|\Psi (0)\ra= |\psi_0^S\ra \otimes|0\ra_{A}\otimes|\varphi_0\ra_{B} $,
 where $|\psi_0^S\ra$ and $|\varphi_0\ra_{B}$
 are vectors in the Hilbert spaces of $S$ and of $B$, and
 $|0\ra_A$ is an eigenstate of $H_A$.
 The quantum kicked rotor is quantized on a phase space torus with a Hilbert space dimension $N=4096$,
 whose initial state is taken as a randomly generated vector from its Hilbert space
 (quantum recurrence time is already sufficiently large).
 The kicking period $T$ is taken
 as $2\pi/N$.  Typical values of $\omega_x$, $\omega_z$, and $\omega_A$ are set around $10^3$ in
 dimensionless units, such
 that within one kicking period $T$ the characteristic phase evolution of the two-level systems
 are of the order of unity.  Many initial states were studied but in Fig.~1 we
only report representative results for one initial state.
    Note also that one
 key parameter is $\varepsilon$, which
 represents the strength of system-environment coupling.

 \begin{figure}
    \includegraphics[width=8.2cm]{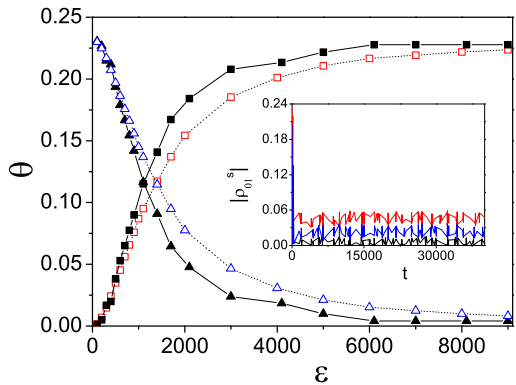}
     \caption{(Color online)
     Angle $\theta$ (in unit of radiant) to be rotated to reach state $|\ww \rho_1\ra$ (empty symbols)
     from one eigenstate of $H_{S}$ (square) or of $H_{I}$ (triangle), for a wide range of
     $\varepsilon$.
     State $|\ww\rho_1\ra $ is one numerically found eigenstate of ${\overline{\rho}}$
     (a time-averaged RDM for $t\in [30000T, 40000T]$). For comparison, the $\theta$ values to
     reach the state $|\alpha\ra $ (filled symbols) determined theoretically
     [details after Eq.~(6)] by maximizing $\| \Delta H \|$, are also presented.
The initial state of the system $S$ is placed in a superposition state
$ {0.8}\exp(5i)|x_{+}\rangle + {0.6}|x_{-}\rangle$, where $|x_+\rangle$ and
$|x_-\rangle$ represent spin-up and spin-down states along the $x$-direction.  For the initial state of
the environment, $A$ is placed in an eigenstate of $H_A$ and the kicked-rotor state is chosen
randomly. Other system parameters are  $\omega_{x}=0.5\times10^3 $, $\omega_{z}=1.0\times10^3$,
   $\omega_{A}=1.5\times10^3$,
       $v=90/T, N=2^{12}$, $\lambda=0.1$.
        Inset: Decay
of the off-diagonal element of RDM with time, in $|E_k\rangle$
representation (upper red curve), in eigenrepresentation
of $H_I$ (middle blue curve), and in representation of the
PS identified here (bottom dotted curve), for $\varepsilon= 2000$.  Note that the decay
is essentially done within about $600 T$.
}
     \label{fig1}
    \end{figure}

As shown in Ref.~\cite{WGCL08}, if the system-environment coupling strength
$\varepsilon$ is below a threshold $\varepsilon_p$, then
the off-diagonal elements of the RDM in the eigenrepresentation of
$H_{S}$ will show a Gaussian-type decay.
The dephasing time of energy eigenstates ($T_2)$ then scales as $\varepsilon^{-1}$,
whereas the population relaxation time $(T_1$) goes as
$\varepsilon^{-2}$ (obtained from Fermi's golden rule). In our model we find $\varepsilon_p\sim 1/N$.
So for $\varepsilon<1/N$, $T_2 \ll T_1$, and hence the energy eigenstates $|E_k\rangle$
form the PS.
Detailed calculations from our present model confirm this and
also reveal something  interesting.
As shown in Fig.~1 (empty squares), $|E_k\rangle$ are found to agree well with the computationally
found PS (i.e., very small values of $\theta$)
for $\varepsilon$ as large as $10^{1}$.
Though our previous work~\cite{WGCL08} did not rule out the possibility
of $|E_k\rangle$ being the PS for $\varepsilon>\varepsilon_p$, it is remarkable
to see that $|E_k\rangle$ here still form the PS even for $\varepsilon \gg \varepsilon_p $.
This should be related to
the fact that here the system ($S$) is only directly coupled with a small component ($A$) of
the environment.

 \begin{figure}
     \includegraphics[width=8.2cm]{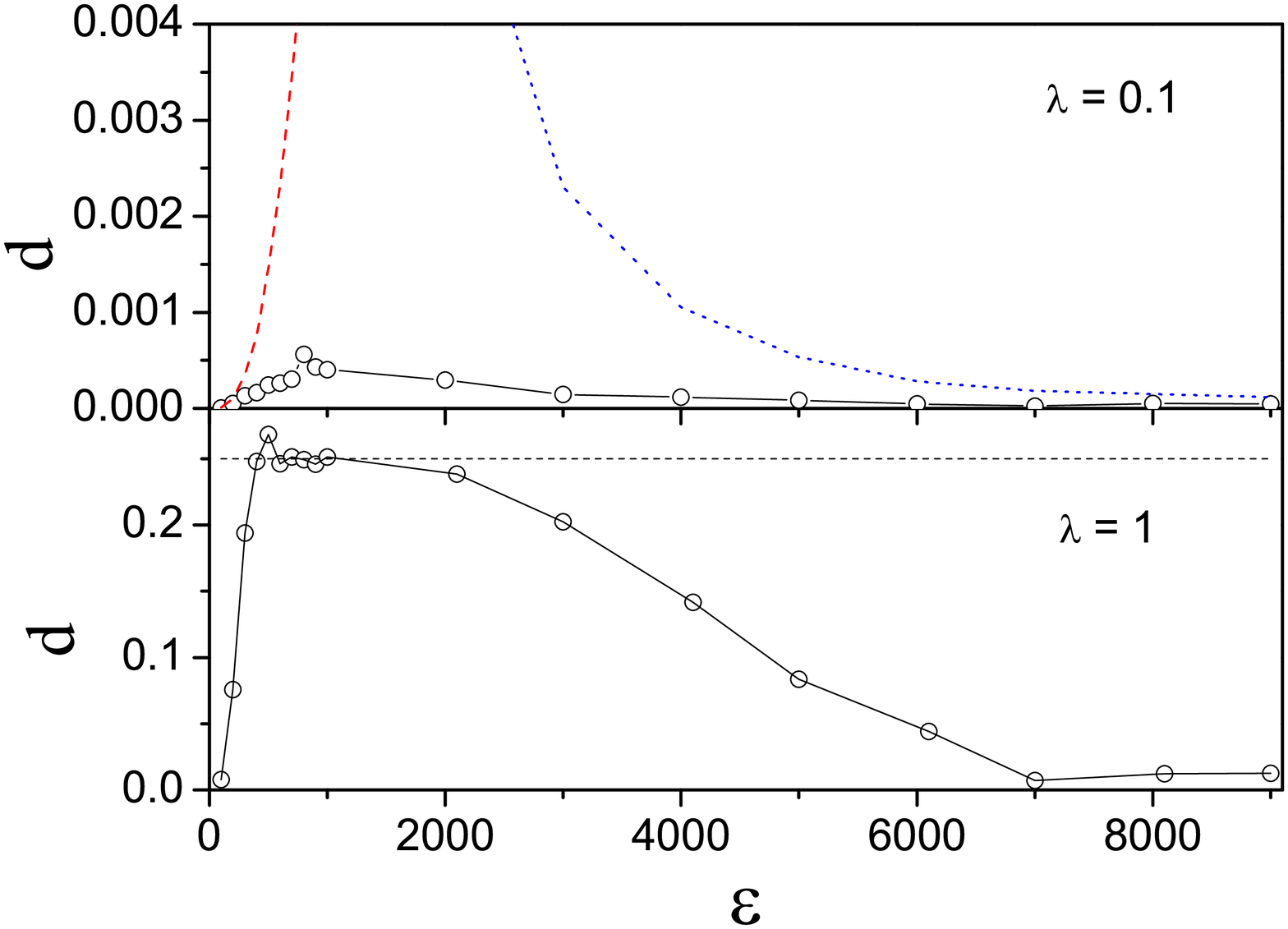}
     \caption{(Color online)
Upper panel:  Behavior of the time-evolving eigenstates of the system's RDM, as described by a
time-averaged distance $d$ from the states
$(|\ww \rho_0\ra, |\ww \rho_1\ra)$ found computationally in Fig.~1 (circles), from eigenstates of
$H_S$ (dashed red line), and from eigenstates of $H_I$ (dotted blue line), for a wide range
      of $\varepsilon$. Other system
     parameters are the same as in Fig.1. Bottom panel: distance $d$ between time-evolving eigenstates
     of the system's RDM and
       states $(|\ww \rho_0\ra, |\ww \rho_1\ra)$ found computationally in the case of $\lambda=1.0$.
       Large $d$ values
       in the bottom panel indicate the loss of PS.
 }
     \label{fig2}%
 \end{figure}

Results for larger values of $\varepsilon$ are also detailed in Fig.~1.  Consider first
$\varepsilon$ values approaching $10^4$, i.e., the right end of the curve shown with empty triangles.
In these cases, the computationally found states $|\ww{\rho}_k\rangle$
are rotated from the
$H_I$ eigenstates by essentially a zero angle.
Hence, the eigenstates of $H_I$ can be regarded as the PS in this strong coupling case, even for
a time scale much larger than the characteristic scale of $H_S$.
Next we turn to intermediate cases with $\varepsilon\in [10^2, 5\times 10^3]$.
As seen from Fig.~1, states $|\ww{\rho}_k\rangle$
can notably deviate from $|E_k\rangle$ as well as the $H_I$ eigenstates.  As we tune up the
value of $\varepsilon$,
states $|\ww{\rho}_k\rangle$ exhibit a clear and smooth transition from being close to
$|E_k\rangle$ to being close to
the $H_I$ eigenstates.  Further, as a consistency check, the inset of Fig.~1 shows the decay
of the off-diagonal elements of the RDM
in three representations, for $\varepsilon=2000$ as an example.  It is seen that only in the
$|\ww{\rho}_k\rangle$ representation, the off-diagonal elements
decay to small values with some fluctuations \cite{note3}.

It is yet to be shown that at sufficiently later times the eigenstates
$|\rho_k(t)\ra$ of the time-evolving RDM
only slightly fluctuate around $|\ww{\rho}_k\rangle$.
The upper panel of Fig. 2 depicts
the distance $d(|\ww\rho_{k'}\ra)$ vs $\varepsilon$ (solid line) (i.e., for the PS identified in Fig.~1).
It is seen that for the entire considered regime of
$\varepsilon$, $d$ remains impressively small. For intermediate values of $\varepsilon$, it is
much smaller than
the same $d$-distance between $|\rho_k(t)\ra$ and eigenstates of $H_S$ (dashed red line) or
between $|\rho_k(t)\ra$ and eigenstates of $H_I$ (dotted blue line).
Therefore, after an initial period the time-evolving RDM eigenstates
$\{|\rho_0(t)\rangle,|\rho_1(t)\rangle\}$  do remain close to $\{|\ww{\rho}_0\rangle,
|\ww{\rho}_1\rangle\}$, suggesting that the
RDM becomes almost diagonal in the $|\ww{\rho}_k\rangle$ representation, even for intermediate
values of $\varepsilon$.
We finally infer that the states $|\ww{\rho}_k\rangle$, whose behavior is shown in Fig.1,
are indeed excellent PS emerging from the coherent quantum dynamics of $S+\E$.
Other detailed calculations also indicate that PS does not always exist. For example,
if the coupling between $A$ and $B$ is also very strong (e.g., $\lambda=1.0$, bottom panel of Fig.~2),
then the diagonal elements of RDM become very close, $d$ can reach as large as $0.25$ (dashed line)
(a value that can be estimated theoretically \cite{epaps}),
and consequently PS is lost.

\emph{Theoretical insights} --- We shall now develop some insights into
our computational results.  We rewrite the total state for $S+\E$ as
\be |\Psi(t)\ra  =  |\alpha \ra |\phi_{\alpha }(t)\ra
 + |\beta \ra |\phi_{\beta }(t)\ra ,
 \label{expand2}
\ee
where $(|\alpha\rangle, |\beta\rangle$) is a chosen time-independent orthonormal basis set
in the Hilbert subspace for $S$,
$|\phi_{\alpha }(t)\ra$ and  $|\phi_{\beta }(t)\ra$ are the associated ``expansion states"
living the Hilbert subspace for $\E$.  The time-dependence
  of the off-diagonal element of the system's RDM is then given by
\bey \rho^s_{\alpha \beta } \equiv \langle\alpha|\rho^{s}|\beta\ra
 =  \la \phi_{\beta}(t)|\phi_{\alpha }(t)\ra . \eey
With this notation, seeking PS then becomes the search for $(|\alpha\rangle, |\beta\rangle$),
such that the evolution of
 $|\phi_{\alpha }(t)\ra$ is as different as possible from $|\phi_{\beta }(t)\ra $.
This insight motivates us to examine the time evolution of the states $|\phi_{\alpha }(t)\ra$ and
$|\phi_{\beta }(t)\ra$
used in Eq.~(\ref{expand2}). To that end, we first define  $H^S_{\alpha \beta}\equiv
\la \alpha |H_S|\beta \ra $, $H^I_{\alpha \beta} \equiv \la \alpha |H_I|\beta \ra $
(and those by $\alpha\leftrightarrow\beta$). Note that
$H^S_{\alpha \beta}$ thus defined is a scalar, but $H^I_{\alpha \beta}$ is still an operator
on the Hilbert subspace for $\E$.  Just to have a rather compact Schr\"{o}dinger-like equation for
$|\phi_{\alpha }(t)\ra$ and $|\phi_{\beta }(t)\ra$, we introduce more
operators on the $\E$ subspace, i.e.,
$H_{\alpha\alpha} \equiv H_\E + H^I_{\alpha \alpha } + H^S_{\alpha\alpha}$, $H_{\alpha \beta}
\equiv H^S_{\alpha \beta } + H^I_{\alpha \beta}$,
$K_\alpha \equiv H_{\alpha \beta}H_{\beta \beta}H_{\alpha \beta}^{-1}$, and $\ J_\alpha \equiv
H_{\alpha \beta}H_{\beta\alpha}$.
Using these definitions and the Schr\"{o}dinger equation
for $S+\E$, we obtain
 \bey \label{phi-ev}
  i\frac{d}{dt} |\phi_{\alpha}\ra = H_{\alpha \alpha}|\phi_\alpha \ra + i |\xi_\alpha \ra ,
  \eey
  where $|\xi_\alpha \ra\equiv -iH_{\alpha\beta}|\phi_{\beta}\rangle$, with
  \bey i\frac{d}{dt}|\xi_\alpha \ra
   =  K_\alpha |\xi_\alpha \ra -iJ_{\alpha}|\phi_\alpha\ra. \label{phi-ev2}
 \eey
An analogous equation for  $|\phi_{\beta }(t)\ra$ is obtained by exchanging $\alpha$ and $\beta$.

Equation~(\ref{phi-ev}) indicates that the difference between the evolution
of $|\phi_{\alpha }(t)\ra$ and that of  $|\phi_{\beta }(t)\ra$ is caused by the
difference between the operators $H_{\alpha\alpha}$ and $H_{\beta\beta}$ and by
the difference between $|\xi_\alpha\rangle$ and $|\xi_{\beta}\rangle$. Further,
Eq.~(\ref{phi-ev2}) shows that $|\xi_\alpha\rangle$ and $|\xi_{\beta}\rangle$ evolve differently
due to the difference between $K_{\alpha}$ and $K_{\beta}$ and between $J_{\alpha}$ and $J_{\beta}$.
To quantify these operator differences we define
$\Delta H \equiv H_{\alpha \alpha}-H_{\beta \beta}$, $\Delta K \equiv K_{\alpha}-K_{\beta}$,
and $\Delta J \equiv J_{\alpha}-J_{\beta}$.  In our model,
because $H_I$ is a direct product of two spin operators, one finds $\Delta J=0$~\cite{epaps}.
Furthermore, for $\varepsilon$ much less or much larger than the energy scale of $H_{S}$, we
find $\Delta K \approx -\Delta H$~\cite{epaps}.  For cases with intermediate values
of $\varepsilon$, a more detailed analysis~\cite{epaps} gives again that
$\Delta K \approx  -\Delta H $, at least for those states $(|\alpha\ra,|\beta\ra)$ that
maximize $||\Delta H||$, where $||\Delta H||$
represents the Frobenius-2 norm, a simple measure of $\Delta H$.
Putting these observations together, we intuitively expect (not a proof)
that for the entire considered regime of $\varepsilon$, the basis states $(|\alpha\ra,|\beta\ra)$
maximizing $||\Delta H||$ may approximately give the most substantial difference
between $|\phi_{\alpha }(t)\ra$ and $|\phi_{\beta }(t)\ra$ and hence the most significant
decay of $|\rho^s_{\alpha \beta}|$. As such, the basis states $(|\alpha\ra,|\beta\ra)$
theoretically determined by maximizing $||\Delta H||$ should agree with the PS computationally obtained above.

In Fig.1 we compare the PS (empty symbols) found from the decoherence
dynamics with the states $(|\alpha\ra,|\beta\ra)$ (filled symbols) determined directly by
maximizing $||\Delta H||$.
In terms of their relation with the eigenstates of $H_S$ and of $H_I$, nice agreement is
obtained for the whole
regime of $\varepsilon$.

It is also interesting to discuss the implication of the term $i|\xi_\alpha \ra$ in Eq.~(\ref{phi-ev}).
Note that this term does not preserve the norm
$\langle\phi_{\alpha}|\phi_{\alpha}\ra$. Hence, populations on the basis states $(|\alpha\ra,|\beta\ra)$,
 even when they are identified as the PS, can still fluctuate with time. This constitutes a
  crucial difference from a pure-dephasing problem.  Unlike in a pure-dephasing problem,
    here the decay of $\rho^s_{\alpha \beta}=\langle\phi_{\beta}(t)|\phi_{\alpha }(t)\rangle$
     cannot be interpreted as that of the overlap of
two independent
environment histories $ |\phi_{\alpha }(t)\ra$ and $ |\phi_{\beta}(t)\ra$. Instead, these
two evolution histories are mingled together through
population transitions between them.  Difference from a pure-dephasing picture is also made evident
by the role of the term $\Delta H^{S}\equiv H^{S}_{\alpha\alpha}-H^{S}_{\beta\beta}$ as one component
of $\Delta H=\Delta H^{S}
+H^{I}_{\alpha\alpha}- H^{I}_{\beta\beta}$.
For a pure-dephasing problem, i.e., if the term $i|\xi_\alpha \ra$ is switched off, then the
 component $\Delta H^{S}$ becomes irrelevant: it is a c-number for the environmental Hilbert subspace,
 and hence cannot cause the decay of $|\langle\phi_{\beta}(t)|\phi_{\alpha}(t)\ra|$.
By contrast, in our model the term $\Delta H^{S}$ in $\Delta H$ is found to be necessary for
 predicting the PS with intermediate system-environment coupling. That is, without this term,
  maximization of $||\Delta H||$ would incorrectly predict that eigenstates of $H_I$ are the
   PS regardless of the value of $\varepsilon$.

{\it Conclusion} ---
The concept of PS may still apply {if an environment with many degrees of freedom (like
a thermal bath) is replaced by an environment with very few degrees of freedom.}
 Approximate PS are shown to exist for intermediate system-environment coupling and can
continuously deform to expected limits.
Such types of PS emerging from quantum dynamics alone
are of importance to understanding decoherence and thermalization processes.

 W.G. was supported by the Natural Science Foundation of China under Grants No.~10775123
 and No.10975123 and the National Fundamental Research Programme of China
 Grant No.2007CB925200.  J.G. was supported by the NUS-``YIA"-R-144-000-195-101.

 \subsection{Supplementary Materials I:
 Some operators and equations used in the manuscript}

 To qualitatively understand the emergence of
 preferred (pointer) states (PS) for intermediate strength of system-environment coupling,
 we first recapitulate some equations and operators defined in our manuscript. In particular,
 the ``expansion states" living in the Hilbert subspace of the environment is denoted by
 $|\phi_{\alpha}(t)\rangle$, with $|\Psi(t)\ra  =  |\alpha \ra |\phi_{\alpha }(t)\ra
  + |\beta \ra |\phi_{\beta }(t)\ra$,
 where $(|\alpha\rangle, |\beta\rangle$) is an time-independent orthonormal basis set
 in the Hilbert subspace of the system $S$.

 Using Schr\"{o}dinger equation for the system and the environment
 as a whole, under the condition that $H_{\alpha \beta}$ does not explicitly depend on the time $t$
 and is a reversible operator,  the equation of motion of $|\phi_{\alpha}(t)\rangle$ is
 given by
 \bey \label{phi-ev}
  i\frac{d}{dt} |\phi_{\alpha}\ra = H_{\alpha \alpha}|\phi_\alpha \ra + i |\xi_\alpha \ra ,
 \\ i\frac{d}{dt}|\xi_\alpha \ra
   =  K_\alpha |\xi_\alpha \ra - iJ_{\alpha}|\phi_\alpha\ra, \label{phi-ev2}
 \eey
 where \bey H_{\alpha\alpha} = H_\E + H^I_{\alpha \alpha } + H^S_{\alpha\alpha}, \nonumber
 \\ K_\alpha = H_{\alpha \beta}H_{\beta \beta}H_{\alpha \beta}^{-1}, \nonumber
 \\ J_\alpha=H_{\alpha \beta}H_{\beta\alpha},
 \eey
 with
 \begin{eqnarray} H_{\alpha \beta} &=& H^S_{\alpha \beta } + H^I_{\alpha \beta}, \nonumber
 \\ H^S_{\alpha \beta}& =& \la \alpha |H_S|\beta \ra, \nonumber
 \\ H^I_{\alpha \beta}& =& \la \alpha |H_I|\beta \ra,
 \end{eqnarray}
 and analogous expressions by exchanging $\alpha$ and $\beta$.
 In the manuscript we then defined the following operators:
 \begin{eqnarray} \Delta H &\equiv & H_{\alpha \alpha}-H_{\beta \beta}, \nonumber \\
 \Delta K &\equiv& K_{\alpha}-K_{\beta}, \nonumber \\
 \Delta J &\equiv& J_{\alpha}-J_{\beta}.
 \end{eqnarray}

 The vector $|\xi_\alpha\ra $ is in fact related to $|\phi_{\beta}\ra $ by
 $|\xi_\alpha\ra = -i H_{\alpha \beta }|\phi_{\beta}\ra $.
 For an initial system-environment direct-product state $|\Psi(0)\ra = |\psi^S_0\ra |\phi_0\ra $,
 we have $|\phi_\alpha(0)\ra = (\langle\alpha|\psi_0^S\rangle) |\phi_0\ra$ and
 $ |\phi_{\beta}(0)\ra = (\langle\beta|\psi_0^S\rangle) |\phi_0\ra $.
 Thus, the initial condition of Eqs.~(\ref{phi-ev}) and (\ref{phi-ev2}) is given by
 $|\phi_\alpha(0)\ra $ and
 $|\xi_\alpha(0)\ra = -i(\langle\beta|\psi_0^S\rangle) H_{\alpha \beta }|\phi_0\ra $.

 Note that $\langle \phi_{\alpha}(t)|\phi_{\alpha}(t)\ra$ and  $\langle \phi_{\beta}(t)|\phi_{\beta}(t)\ra$
 give the probabilities of finding the system in the states $|\alpha\ra$ and $|\beta\ra$,
 respectively. During the time evolution, these two quantities need not to be conserved because
 the problem of interest here is not a pure-dephasing problem.

 \subsection{Supplementary Materials II:
 On conditions for PS to be identified computationally}

 In our manuscript we also mentioned conditions under which we can computationally
  identify PS from the time-evolving reduced-density matrix (RDM) of the system. These conditions
  are further elaborated in this section.
 After some simple derivation, we obtain the following expression for the distance
 $D(|\rho_{k}(t)\ra,|\alpha\ra ) = 1- |\la \rho_k(t)|\alpha\ra |^2$, where
 the eigenstate $|\rho_k(t)\rangle$ of the RDM $\rho^{s}(t)$ is chosen by the condition
  $|\langle\rho_k(t)|\alpha\rangle|^2 \ge 1/2$,
 \bey \label{ov-ge} D(|\rho_{k}(t)\ra,|\alpha\ra ) = \frac{|\rho^s_{\alpha \beta}|^2}
     {\frac{1}{4}\left[\delta\rho + \sqrt{(\delta\rho)^2+4|\rho^s_{\alpha \beta}|^2}
     \right]^2    +|\rho^s_{\alpha \beta}|^2}. \ \
 \eey
 Here, $\delta \rho \equiv |\rho^s_{\alpha \alpha} - \rho^s_{\beta \beta}|$, with
 $\rho^{s}_{\alpha\alpha}$ ($\rho^{s}_{\alpha\beta}$) being  the
 diagonal (off-diagonal) matrix elements of the RDM.   The time-averaged value of
 $D(|\rho_{k}(t)\ra,|\alpha\ra )$ is denoted by $d(|\alpha\ra )$ in the manuscript.

 Suppose at long times, the overlap $\la \phi_{\beta}(t) |\phi_{\alpha}(t)\ra $
 is already of a small magnitude.
 Then, $\rho^s_{\alpha \beta}=\la \phi_{\beta}(t)|\phi_{\alpha}(t)\ra $ is small by definition.
 Furthermore, we assume that the populations in the two basis states, namely, $\rho^s_{\alpha \alpha}$ and
 $\rho^s_{\beta \beta}$, have also approached some equilibrium values at long times and can maintain an appreciable difference
 $\delta \rho$ such that $\delta\rho \gg |\rho^s_{\alpha \alpha}|$.
 Then, Eq.~(\ref{ov-ge}) becomes
 \bey \label{ov-sp}
   D(|\rho_{k}(t)\ra,|\alpha\ra ) \simeq \frac{|\rho^s_{\alpha \beta}|^2}
     {(\delta\rho)^2 }.
 \eey
 This then gives
 \be \label{d-ex} d(|\alpha\ra ) \simeq \left[ \frac{1}{t_{b}-t_{a}}\int_{t_a}^{t_b}|\rho^s_{\alpha \beta}|^2 \ dt\right]  \frac{1} {(\delta \rho)^2} \ll 1 . \ee
 Sufficiently small value of $d(|\alpha\ra )$
 indicates that eigenstates of the RDM $\rho^s(t)$ only have small fluctuations around
 the fixed set of basis $(|\alpha \ra , |\beta\ra )$.
 This fixed set of basis states then form the PS, at least approximately.
In our manuscript we computationally identified such a basis set, which is denoted by
$|\tilde \rho_k\ra$.

It is now clear that PS can be computationally identified under the following conditions:
 (1) small $|\rho^s_{\alpha \beta}|$ and (2) relatively large $\delta \rho$.
 Smallness of $|\rho^s_{\alpha \beta}|$ is equivalent to the smallness of the overlap
 $|\la \phi_{\beta}(t)|\phi_\alpha (t)\ra|$. This occurs if
 $\Delta H$, $\Delta K$, and $\Delta J$ are made as different as possible, because these operators will
 determine the evolution difference between $|\phi_{\beta}(t)\rangle$ and $|\phi_{\alpha}(t)\rangle$.

 \subsection{Supplementary Materials III:Distance $d$ averaged over the Hilbert space}

In the case of the bottom panel of Fig.2, the environment part $A$ is also strongly coupled with the kicked rotor,
the distance $d$ is large and hence the PS is lost.  Here we give a simple explanation.
Due to the strong coupling within the environment and the strong coupling between $S$ and $\E$,
we can assume that the motion is sufficiently random in the total Hilbert space for $S+\E$. Therefore,
the long-time-averaged $d$ can be approximated by a Hilbert space average. Let $\langle d \rangle$ denote this total-Hilbert-space-averaged
distance describing the difference between RDM eigentates and a fixed set of basis states
 $(|\alpha\ra, |\beta\ra)$.  In the following, we estimate $\la d\ra$ using a method similar to the canonical
typicality approach  discussed in Ref.~\cite{Gold06}.
The dimension of the Hilbert subspace of the environment $\E$ is denoted by $n_{\E}$ and the dimension of $S+\E$ is $2n_{\E}$.

 Assuming that the dynamics is almost ergodic on the total Hilbert space, a ``typical" vector in the total Hilbert space
 can be written as
\bey \label{Psi} |\Psi\ra = \N \sum_\alpha |\alpha\ra |\Phi^{\E}_\alpha\ra ,
 \ \ \text{with} \ \label{Phi} |\Phi^{\E}_\alpha\ra  = \sum_{j} C_{\alpha j} |j\ra ,
 \eey
 where $\{ |j\ra \}$ is an orthonormal basis set in the Hilbert subspace for the environment $\E$,
 $\N$ is the normalization coefficient,
 \be \N = \frac{1}{\sqrt{\sum_{\alpha j} |C_{\alpha j}|^2}} \simeq \frac{1}{2n_{\E}}, \ee
 and the real and imaginary parts of $C_{\alpha j}$ are assumed to be independent real Gaussian random
 variables with mean zero and variance 0.5.  For this typical vector,
the elements of the RDM are given by
 \bey \label{rhos1}
     \rho^s_{\alpha\beta} & = & \N^2 \sum_{j}C_{\beta
     j}^{\ast}C_{\alpha j},
     \\    \delta \rho & = & |\rho^s_{\alpha\alpha}-\rho^s_{\beta \beta}| = \N^2
     \left |  \sum_{j} |C_{\alpha j}|^2-|C_{\beta j}|^2 \right | . \ \
 \label{drho1}  \eey
 In the limit of large $n_\E$, after simple derivation one finds
 \bey \lim_{n_\E \to \infty} \frac { \la |\rho^s_{\alpha\beta}|^2 \ra }{ \la \delta \rho^2 \ra } = \frac{1}{2},
 \label{ra} \eey
 with the average $\langle\cdot\rangle$  taken over all typical vectors defined above.
 Furthermore, using Eq.~(\ref{ov-ge}),
 one obtains that $\la d \ra =0.25$, a result consistent with the bottom panel of Fig.~2. This clearly shows
 that PS does not always exist.


 \subsection{Supplementary Materials IV: Analysis of $\Delta H$, $\Delta K$, and $\Delta J$}

 Finally, we discuss how the three difference operators $\Delta H$, $\Delta K$, and $\Delta J$
 may be used to predict PS without actually following the decoherence dynamics computationally.  The qualitative picture is
 that in the PS representation, the operators $\Delta H$, $\Delta K$, and $\Delta J$ should be maximized such that
 the two evolution histories $|\phi_{\beta}(t)\rangle$ and $|\phi_{\alpha}(t)\rangle$ are as different as possible and hence the off-diagonal elements of
 the RDM will decay to very small values.
 Because in our model, $H_I$ has a product form, namely, $H_I = \varepsilon \sigma_z^S \otimes \sigma_z^{A}$, we have
 \begin{eqnarray}
  [H^I_{\beta \alpha} , H^I_{\alpha \beta}] &=&0, \\
 \Delta J = H_{\alpha \beta}H_{\beta\alpha} - H_{\beta \alpha }H_{\alpha \beta } &=&0. \end{eqnarray}
 Given $\Delta J=0$, we focus only on the maximization of $\Delta H$ and $\Delta K$ to understand the emergence of PS.

 Interestingly, the two operators $\Delta H$ and $\Delta K$ are also related to each other.  Let $X_\alpha = [H_{\alpha \beta }, H_{\beta \beta}] $, we find (after some straightforward calculations) the following expression
 \be \Delta K = -\Delta H + Y,
 \ee
 where
 \be Y = X_\alpha H_{\alpha \beta}^{-1} - X_{\beta} H_{\beta \alpha }^{-1}.
 \ee

For our model, we may further simplify $X_{\alpha}$ and obtain
 $X_{\alpha }  
  = [ H^I_{\alpha \beta}, H_{\E}] $.
Next, to have explicit expressions of operators, we now take the eigenstates of $\sigma_x$ as our working
representation, where $|1\ra_{x}$ and $|0\ra_{x}$ represent spin up and down along the $x$-axis. We expand a given set of basis states $\{|\alpha\ra,|\beta\ra\}$
 as the following:
 \begin{eqnarray}
  |\alpha\ra&=&ae^{i\varphi}|1\ra_{x} + b|0\ra_{x}, \nonumber \\
   |\beta\ra&=&be^{i\varphi}|1\ra_x
 - a|0\ra_{x}\ ,\end{eqnarray}
 where $a,b\in [0,1]$ are real expansion coefficients.
 Using these two explicit expressions and our direct product form of $H_I$, we find
 \bey
 X_\alpha 
    =  [(1-2b^2)\sin (\varphi)+i\cos (\varphi)](2i \varepsilon \omega_A)\sigma_y^A
\eey
 and
 \bey Y 
               = 2i \varepsilon \omega_A(\eta L - \ov\eta \ov L) ,
 \eey
 where $\sigma_y^{A}$ is the Pauli operator along the $y$-axis for the two-level system A
 as a part of the environment,
 \bey
   \eta = \frac{c_1}{d_{1}^{2}-\varepsilon^2 c_{1}^{2}}\mbox{\ ,\ }
   \ov\eta = \frac{c_2}{d_{2}^{2}-\varepsilon^2 c_{2}^{2}},
   \nonumber \\
   c_1 = (1-2b^2)\sin(\varphi)+i\cos(\varphi), \nonumber\\
   d_1 = 2\omega_{x} \sqrt{1-b^2}b + \omega_{z}c_1, \nonumber
  \nonumber \\  c_2=c_{1}^{\ast}, \ \    d_2=d_{1}^{\ast},
   \eey
   and
 \bey
    L=-i\varepsilon c_1\sigma_{x}^A+ d_1\sigma_{y}^A,
    \ov L=-i\varepsilon c_2\sigma_{x}^A+ d_2\sigma_{y}^A.
 \eey

 \begin{figure}
     \includegraphics[width=8.0cm]{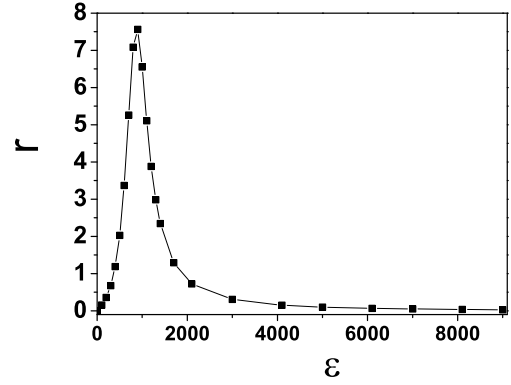}
     \caption{ Variation of the ratio $r$ with $\varepsilon$ for an arbitrarily chosen
     basis $(|\alpha\ra , |\beta\ra )$.
     Parameters: the same as in Fig.1 of the text.
 }
     \label{fig1}%
 \end{figure}

Let us consider the first limiting case of $\varepsilon \ll \omega_x$. Neglecting terms of the order of $\varepsilon / \omega_x$
 and noticing that $\omega_i$ ($i=x,z,A$) are of the same order of magnitude, we find
 \bey
   \eta\simeq \frac{c_1}{d_1^{2}},\ \ \ \ov\eta\simeq
   \frac{c_2}{d_2^{2}}\nonumber\\
   L\simeq d_1 \sigma_y^A, \ \ \ \ov L \simeq d_2 \sigma_y^A .
 \eey
 Thus,
 \bey
   Y \simeq 2i\varepsilon
  \left ( \frac{\omega_{A}c_1}{d_1}-\frac{\omega_{A}c_2}{d_2} \right ) \sigma_y^A \sim \varepsilon .
 \eey
Clearly  then, because $\Delta H \sim \omega_x $, one infers that $\Delta K \simeq -\Delta H $.

In the second limiting case of $\varepsilon \gg \omega_x$, we neglect all terms of the order of
 $\omega_x/\varepsilon$ and then obtain
 \bey
   \eta\simeq -\frac{1}{\varepsilon^{2}c_1}, \ \ \ov\eta\simeq
   -\frac{1}{\varepsilon^{2}c_2}\nonumber\\
   L\simeq -i\varepsilon c_{1} \sigma_x^A,\ \ov L \simeq -i\varepsilon c_{2} \sigma_x^A.
 \eey
 Thus,
 \bey   Y \simeq 2 \omega_A \sigma_x^A -  2 \omega_A \sigma_x^A = 0.
 \eey
 This also leads us to the expectation that $\Delta K \simeq -\Delta H $.  That is, for very small or very large system-environment coupling strength $\varepsilon$, we always have $\Delta K \simeq -\Delta H $.  %

 The situation of intermediate system-environment coupling strength with
 $\varepsilon \sim \omega_x$ is more complicated.  To better understand this case we have
  performed numerical
 calculations of the ratio,
 \be r (\varepsilon, |\alpha\ra ) = \frac{ \| Y \| }{\| \Delta H \| }, \ee
 where $\| \ldots \|$ indicates a Frobenius-2 norm.
 Figure 1 shows the dependence of the ratio $r$ on $\varepsilon$ for an arbitrarily chosen basis
 $(|\alpha \ra , |\beta\ra )$. It shows that
 the ratio can be large for intermediate values of $\varepsilon$.
 Therefore, $\Delta K$ may have considerable deviations from $-\Delta H$
 for a general basis set $(|\alpha \ra , |\beta\ra )$.
 On the other hand, for the basis set that maximizes the norm $\| \Delta H \|$, it is found that
 the ratio $r$ is negligibly small, as shown in Fig.2.

 \begin{figure}
     \includegraphics[width=8.0cm]{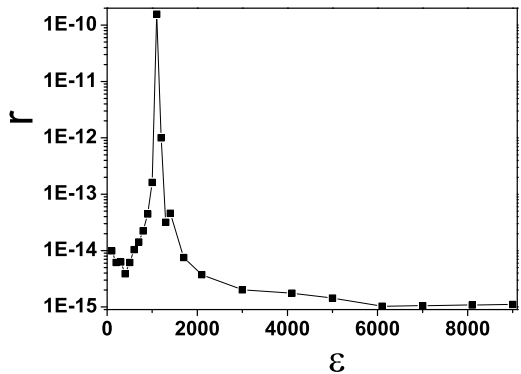}
     \caption{ Variation of the ratio $r$ with $\varepsilon$ for the basis $(|\alpha\ra , |\beta\ra )$
     that maximizes the norm $\| \Delta H \|$.
 }
     \label{fig2}%
 \end{figure}

 Summarizing the above considerations, we have that
 for weak, strong, and intermediate system-environment coupling strength, maximizing
 $\|\Delta H\|$ is expected to maximize the $\|\Delta K\|$, and as a result, the properties of PS
 can be connected with a maximization of $\|\Delta H\|$.
 Therefore, so long as the impact of $\Delta H$ and $\Delta K$ can be quantitatively captured by their
 norms $\| \Delta H \|$ and $\| \Delta K \|$, we may consider $\|\Delta H\|$ only.
 In particular, by maximizing $\| \Delta H \|$, we expect that the two states
 $|\phi_{\alpha}(t)\rangle$ and $|\phi_{\beta}(t)\rangle$ in Eqs.~(\ref{phi-ev}) and
 (\ref{phi-ev2}) will evolve most differently, thus yielding an approximate method of predicting and
 understanding the emergence of PS.

As also stressed in the manuscript,
 since $\Delta H$ is important in both equations of (\ref{phi-ev}) and (\ref{phi-ev2})
 [$-\Delta K$ contains $\Delta H$ in Eq.(\ref{phi-ev2})],  the contribution of
 the c-number part of $\Delta H$ (i.e., $H^{S}_{\alpha\alpha}-H^{S}_{\beta\beta}$) is nontrivial in
 determining the PS.
 Without this c-number term, maximization of $\Delta H$ would incorrectly predict that eigenstates
 of $H_I$ are the PS regardless of the value of $\varepsilon$.
  This interesting feature is much different from
 a pure-dephasing problem. Indeed, in a pure-dephasing problem, there would be no population
 transitions between $|\phi_{\alpha}(t)\rangle$ and $|\phi_{\beta}(t)\rangle$, and hence this
 c-number term in a pure-dephasing problem would be of no interest: it only induces a phase
 difference between $|\phi_{\beta }(t)\ra$ and $|\phi_{\beta}(t)\ra$ and therefore it does not cause the overlap $\langle \phi_{\beta }(t)|\phi_{\alpha}(t)\ra$ to decay.

As two consistency checks, we first return to the first limiting case of $\omega_x \gg \varepsilon$. In this
case
 \be \label{DK-HS} \Delta H \simeq H^S_{\alpha\alpha} - H^S_{\beta \beta }. \ee
 Because eigenstates of $H_S$ correspond to the maximum of
 $|H^S_{\alpha\alpha} - H^S_{\beta \beta }|$, hence, a maximization of $\| \Delta H \|$,
the energy eigenstates of $H_S$ should be the PS. In the second limiting case
$\omega_x \ll \varepsilon$, we have
 \be \Delta H \simeq \epsilon \left ( \la \alpha |\sigma^S_z |\alpha\ra -  \la \beta |\sigma^S_z |\beta \ra
 \right ) \sigma^{A}_z. \ee
 Therefore, the eigenstates of $\sigma_z^S$ and hence the eigenstates of the interaction Hamiltonian $H_I$,
 can maximize $\|\Delta H\|$, which is again in agreement with the known result that eigenstates of
 $H_I$ are the PS in this strong coupling case.


\begin{thebibliography}{99}
\bibitem{epaps}See supplemenatry material accompanying this submission
 for discussions on conditions for PS to be identified computationally,
 for a distance $d$ obtained by averaging over the Hilbert space, and for a detailed
 analysis in properties of $\Delta H$, $\Delta K$, and $\Delta J$ in the
 model studied in this paper.
\bibitem{notes}Certainly this convention is just one possibility for picking out a good
reference basis set $\{|\ww{\rho}_{0}\rangle , |\ww{\rho}_{1}\rangle\}$
to which the time-evolving states $|\rho_k(t)\ra$ will be compared.
What really matters here is whether PS can be found, i.e., whether the eigenstates of the
RDM can indeed stay close to some time-independent basis states.
\bibitem{note}If $\omega_z=0$, then $H_S$ is quantized along $x$  and
$H_I$ is quantized along the perpendicular direction $z$.  For this case,
it was found that fluctuations around the computed intermediate PS are much larger due to close
diagonal elements of the reduced density matrix.
\bibitem{note3} Because both pointer states were found to have appreciable populations
(with populations around 0.4 and 0.6 in the case of the inset of Fig.\ref{fig1}),
they are not at all fixed-point solutions of the dynamics.
\end{thebibliography}
\end{document}